# Current-Induced Magnetization Control in Insulating Ferrimagnetic Garnets

Can Onur Avci

*Institut de Ciència de Materials de Barcelona (ICMAB-CSIC), Campus de la UAB, Bellaterra, 08193, Spain*

**The research into insulating ferrimagnetic garnets has gained enormous momentum in the past decade. This is partly due to the improvement in the techniques to grow high-quality ultrathin films with desirable properties and the advances in understanding the spin transport within the ferrimagnetic garnets and through their interfaces with conducting materials. In recent years, we have seen remarkable progress in controlling the magnetization state of ferrimagnetic garnets by electrical means in suitable heterostructures and device architectures. These advances have readily placed ferrimagnetic garnets in a favorable position for the future development of insulating spintronic concepts. The purpose of this article is to review recent experimental results of the current-induced magnetization control and associated phenomena in ferrimagnetic garnets, as well as to discuss future directions in this rapidly evolving area of spintronics.**

# 1. Introduction

The discovery of ferrimagnetic garnets (FMGs) dates back to 1950s[1,2] followed by several decades of extensive investigations to understand their structural and magnetic properties.[3] In the early 1970s, the development of liquid phase epitaxy has allowed the growth of micrometer-thick FMG films such as yttrium iron garnet ($Y_3Fe_5O_{12}$, YIG) with ultrahigh quality and low production cost.[4] Consequently, FMGs have become a model system for engineering magnetic and magneto-optical properties with high precision through their composition. FMGs with suitable domain properties and large magneto-optic response have led to technological developments such as magnetic bubble memory,[5] magnetic bubble display,[6] and magneto-optic printers.[7] These advances marked the first integration of magnetic materials into modern electronic technologies in the late 1970s and early 1980s.

The 1980s have marked a paradigm shift in magnetism research. The discovery of giant magnetoresistance has initiated the *spintronics revolution* and boosted the magnetic recording industry.[8,9] For the next two decades, other prominent phenomena such as interfacial perpendicular magnetic anisotropy[10] (PMA) and spin-transfer torques[11-13] (STTs) have spearheaded spintronics research and related applications.[14] All these effects exclusively required conducting ferromagnetic structures. During this period, FMGs have been set aside because of their insulating character, and metallic multilayers have become the preferential materials in this exciting field.

Magnetic insulators, and in particular FMGs, possess extraordinary properties that could be beneficial in spintronics. They are often characterized by low damping and long magnon diffusion lengths, enabling efficient spin current generation and its nonlocal transport.[15-19] They can now be grown in nanometer form, in a highly ordered single-crystal structure, desirable for low-pinning domain wall[20-26] and skyrmion motion.[27,28] They offer multiple degrees of freedom for tuning relevant magnetic properties such as saturation magnetization, magnetocrystalline anisotropy, etc., through composition and strain engineering.[3,29,30] They are naturally protected and highly robust against heat, oxidation, aging, and degradation. Finally, the recently discovered transport phenomena such as spin pumping,[31,32] spin Seebeck effect,[33] and spin Hall magnetoresistance[34] suggest FMGs as highly efficient spin current generating and detecting materials.

Recently, a better understanding of electrical transport in bulk and interfaces of certain material systems characterized by large spin-orbit coupling revealed that relativistic spin-dependent effects could generate pure spin currents as a result of a charge current.[35] The resulting spin-orbit torques (SOTs) can lead to magnetization switching,[36-40] domain wall (DW) motion,[21,22,24,26] magnon generation/suppression,[41] etc., in suitable magnetic heterostructures. In the SOTs scheme, in contrast to the STTs, charge and spin currents travel orthogonally, which provides a simple means to generate spin-torques, among others, on FMGs neighboring to a SOT-source, usually a nonmagnetic metal (NM).

The intersection of FMGs and the SOT-driven magnetization control have created fertile grounds in spintronics. We have seen some remarkable results of SOT-induced magnetization control in FMGs in less than half a decade. In this Review, we will give an overview of some of the pioneering experiments on the SOTs-driven control of FMGs. We will first discuss the characteristics of FMGs, SOTs and other relevant physical phenomena. We will then review some key results and discuss them in a broader context. We will finally discuss the potential directions and future opportunities in this rapidly evolving and exciting area of spintronics.

## 2. General Properties of Ferrimagnetic Garnets

### 2.1 Composition and structure

FMGs constitute a large family of magnetic oxides with the general formula of $X_3Fe_5O_{12}$, where X is either a non-magnetic rare-earth element such as yttrium (Y) or a magnetic rare-earth element such as gadolinium (Gd), terbium (Tb), thulium (Tm), etc. The lattice structure consists of two octahedral $Fe^{3+}$ ions, three tetrahedral $Fe^{3+}$ ions, and three $X^{3+}$ ions at the dodecahedral sites, per formula unit. The octahedral and tetrahedral $Fe^{3+}$ ions are coupled antiparallel to each other while $X^{3+}$, if magnetic, is aligned parallel to the octahedral $Fe^{3+}$. All magnetic interactions are mediated via superexchange interaction through $O^{2-}$ anions. In the case of YIG, there is no magnetic compensation, and it preserves ferrimagnetic order up to 559 K. In iron garnets with magnetic rare-earth elements, the total magnetic moment depends sensitively on the rare-earth; hence the magnetism differs significantly depending on the rare-earth choice, composition, and temperature. Notably, there exists a temperature at which the opposing magnetic sublattices are equal to each other and the film shows zero net magnetization, thereby acts like a collinear antiferromagnet. Nonetheless, the ordering temperature is mainly determined by the magnetic moment of Fe, independent of the rare-earth choice, and hence is about the same (530-580 K) in all FMGs,[42] which makes the entire family of FMGs suitable for room temperature research and applications

*2.2 Growth*

Ultrathin films of FMGs are typically grown on lattice-matched gadolinium gallium garnet ($Gd_3Ga_5O_{12}$, GGG) or similar garnet substrates by pulsed laser or magnetron sputter deposition using stoichiometric targets.[43-45] For the sputter deposition, an off-axis method (i.e., the target and substrate are held at 90° with respect to each other) is generally preferred due to the enhanced stoichiometry of the deposited films.[46] Typically, the substrate is kept at a high temperature (650-850°C) during deposition and slowly cooled down to room temperature to obtain the desired high crystallinity, homogeneity and surface smoothness. Alternatively, room temperature deposition and post-growth annealing are also reported to yield similar results.[47] By employing the mentioned methods, films of thicknesses ranging from a few to tens of nm can be obtained with sub-nm surface roughness.

*2.3 Perpendicular magnetic anisotropy*

Of particular interest to this Review, some FMGs possess bulk PMA, which is an essential property in spintronic devices since it enhances the thermal stability of magnetic elements and ensures long-term data retention. The PMA in FMGs is mainly of magneto-elastic origin, mediated by lattice strain. When grown epitaxially, a small lattice mismatch between the substrate and the FMG film leads to compressive or tensile strain, giving rise to negative magneto-elastic energy and consequently to PMA.[48] Thus far, PMA has been reported for thin films of TmIG,[49-52] TbIG[53], EuIG,[52,53] GdIG,[54] YIG,[55] and Bi substituted YIG[18] (Bi:YIG) grown on GGG or similar substrates, polycrystalline EuIG grown on a quartz substrate,[30] and DyIG grown on a $SiO_2$ substrate by pulsed laser deposition.[56] There were also successful efforts to grow TmIG and YIG films by r.f. magnetron sputtering with similar magnetic and structural properties compared to those obtained by pulsed laser deposition.[46,47,57] Recent studies suggest that there is an additional interface contribution to the PMA from the metal capping of FMGs.[58] For the complementary metal-oxide-semiconductor (CMOS) technology integration, magnetron sputtering is the preferred method since films with high uniformity can be grown on large-scale wafers. Tunable PMA through material and interface engineering and the use of magnetron sputtering together make FMGs very attractive for spintronics.

# 3. Spin-Orbit Torques and Electrical Detection of Magnetization in Ferrimagnetic Garnets

SOTs are presently the state-of-the-art current-induced magnetic manipulation method in spintronic devices, including the ones based on FMGs.[35] Their experimental discovery dates back to the end of the 2000s and early 2010s, first in semiconducting heterostructure,[59] and then in metallic multilayers.[36,60] Initially, the SOTs research has mostly focused on heavy metal/ferromagnet bilayers such as Pt/Co,[36,37,60,61] Ta/CofeB[38,40,62] and W/CoFeB[63] due to their historical significance, technological relevance, the convenience of deposition and fabrication, and large torque-to-current ratios reported. Later on, topological insulators,[64,65] semimetals,[66,67] oxidized metals[68-70] and intermetallic alloys[71,72] combined with appropriate magnetic materials, have been considered. Many SOTs generating mechanisms have been discovered thus far, including the spin Hall effect[63] (SHE), interfacial Rashba-Edelstein effect,[60] topological surface states,[73] bulk crystal asymmetry,[74] anomalous Hall effect,[75] (AHE), planar Hall effect[76] (PHE), interface spin filtering,[77] etc. Among these, the SHE is primarily considered as a SOT source for current-induced magnetization manipulation in FMGs. The SHE emerges in the bulk of the materials characterized by strong spin-orbit coupling (Pt, W, Ta, etc.). It describes the spin-dependent scattering of conduction electrons which generates a pure spin current transverse to the charge current injection direction. A magnetic layer in interfacial contact with a SHE material can then absorb the resulting spin current, which acts as a spin torque on the magnetization and enables its reversal (Fig. 1). In the other listed SOT sources, while the spin current generation mechanism differs, the applied torque has comparable symmetry.

*3.1 Symmetry and characterization of spin-orbit torques*

SOTs consist of damping-like and field-like components with the symmetries $\tau_{DL} \propto \mathbf{m} \times (\mathbf{y} \times \mathbf{m})$ and $\tau_{FL} \propto \mathbf{m} \times \mathbf{y}$, respectively (see Fig. 2a). Here $\mathbf{m}$ is the magnetization unit vector and $\mathbf{y}$ is the in-plane axis perpendicular to the current flow direction. $\tau_{DL}$ has the suitable geometry to induce 180° switching of the in-plane magnetization along the axis perpendicular to the current injection as well as the perpendicular magnetization; though in the latter, an in-plane magnetic field is required to break the rotational symmetry of $\tau_{DL}$ effective field (Fig. 2a). Furthermore, $\tau_{DL}$ can move homochiral Néel-type DWs and skyrmions in the PMA systems with high efficiency and large velocities.

A handful of electrical and optical techniques have been developed to characterize the SOTs.[78] Perhaps the simplest and most accessible method is the harmonic Hall voltage detection of the SOT effective fields.[61,79] It consists of applying an a.c. current through Hall bar devices and detecting and analyzing the harmonic Hall voltage response proportional to **m** (Fig. 2b). The Hall voltage in metallic magnetic systems predominantly consists of AHE and PHE where the former (latter) is proportional to the out-of-plane (in-plane) component of the magnetization vector: $V_\mathrm{H} = R_\mathrm{AHE} I \cos\theta + R_\mathrm{PHE} I \sin^2\theta \sin 2\varphi$. Here $\theta$ and $\varphi$ represent the polar and azimuthal angle of **m** and $I$ is the injected current (see Fig. 3a). Upon an a.c. current injection, **m** oscillates about its equilibrium position due to oscillating SOTs and generates a harmonic response in the Hall voltage, twice the frequency of the input current, hence generating a second harmonic Hall voltage ($V_{2\omega}$). $V_{2\omega}$ contains contributions from SOTs but also thermoelectric effects (anomalous Nernst and spin Seebeck effects) driven by Joule heating-induced temperature gradients.[80] Self-consistent analysis of the data taken in suitable experimental geometries, detailed in Refs.[61,79,80], leads to vector quantification of SOTs in both in-plane and PMA systems, and the proper separation of the non-SOT originating signals.[80,81]

*3.2 Spin Hall magnetoresistance in magnetic insulators*

A charge current cannot pass through FMGs henceforth cannot generate AHE and PHE. However, it has been discovered that FMGs, and in general magnetic insulators, can influence the electrical resistance of the material that is in interfacial contact with it and give rise to spin Hall magnetoresistance[34,82-84] (SMR). SMR, first discovered in YIG/Pt bilayers, describes the changes in the electrical resistivity of a nonmagnetic metal adjacent to a magnetic insulator depending on the magnetization orientation. The effect arises due to the asymmetry in the scattering of the SHE-induced spin current at the FMG/NM interface, which depends on the relative direction of the magnetization with respect to the spin polarization (collinear or orthogonal). The back-scattered spin current contributes to the charge current via the inverse SHE, which ultimately manifests itself as a modification in the longitudinal and transverse resistivity. In a Hall voltage measurement, like the AHE and PHE in conducting magnets, the SMR gives rise to signals proportional to the in-plane and out-of-plane components of the magnetization vector due to the real and imaginary part of the interface spin mixing conductance,[85] respectively. The Hall voltage then reads:

$$V_\mathrm{H} = R_\mathrm{PHE-SMR} I \sin^2\theta \sin 2\varphi + R_\mathrm{AHE-SMR} I \cos\theta + R_\mathrm{OHE} I H_\mathrm{z} \qquad (1).$$

Note, however, that the amplitude of the $R_{\text{PHE-SMR}}$ is generally much larger than $R_{\text{AHE-SMR}}$, contrary to the situation in the metallic systems. Moreover, there is a non-negligible ordinary Hall effect signal ($R_{\text{OHE}}$) linearly proportional to the out-of-plane component of the external field ($H_z$), but unrelated to the magnetization vector, which is generally neglected in all-metal systems due to much larger AHE and PHE contributions. Figure 3 shows a typical characterization of the SMR and ordinary Hall effect signals using a Hall bar device made of TmIG/Pt.[86]

The harmonic Hall voltage method can be used also for quantifying the SOTs in insulating magnets, e.g., in FMG/Pt. Due to the dominant $R_{\text{PHE-SMR}}$ contribution, the in-plane oscillations of **m** induced by SOTs produce the largest $V_{2\omega}$, henceforth, in FMG/Pt the measurement geometry differs from all-metal systems. Furthermore, thermoelectric contributions to $V_{2\omega}$ should be treated with utmost care. In metallic systems, the main thermoelectric contribution is found to originate from the anomalous Nernst effect produced within the magnetic layer itself, whereas in magnetic insulators the spin Seebeck effect (SSE), the subject of the next section, gives the dominant thermoelectric contribution to $V_{2\omega}$. The detailed description of how to extract the damping-like and field-like SOTs and thermoelectric signals in FMG/NM systems with PMA is given in Refs.[87,88]

*3.3 Spin Seebeck effect*

The SSE describes spin current generation by applying a temperature gradient ($\nabla T$) across a magnetically ordered material[33]. In a simplified picture, the temperature difference along the gradient direction creates an imbalance in the magnon population, generating a pure spin current flow along $\nabla T$ with polarization parallel to **m**. The SSE-originated spin current can diffuse into an adjacent metal with a large SHE and generate an inverse SHE voltage or alternatively called a spin Seebeck voltage ($V_{\text{SSE}}$). Additional to the SMR, the in-plane magnetization vector can be probed by measuring $V_{\text{SSE}}$ in FMGs by simply using current-induced Joule heating as a means to generate $\nabla T$ perpendicular to the layers.[89,90] In contrast to the SMR, the SSE can be used to sense 180° rotation of **m** since $V_{\text{SSE}}$ is expected to change sign upon the reversal of **m**. With an a.c. current injection, the SMR and SSE together can be measured by $V_\omega$ and $V_{2\omega}$, respectively, for effectively characterizing the magnetization vector in the 3D space and other relevant properties such as coercivity ($H_c$), magnetic anisotropy ($H_K$), etc. (see Fig. 4). We note that, in addition to the SMR and SSE, it is also possible to detect the magnetization vector orientation in FMGs through spin-torque ferromagnetic resonance measurements[91-93]. However, such measurements rely on a different and more complex experimental setup and signal analysis, hence not depicted in Fig. 4.

**4. Spin-Orbit Torque Switching in Ferrimagnetic Garnets**

Over the past decade, the characterization of various spin-dependent transport phenomena such as SSE, spin pumping, and SMR has shown a significant spin current transmission through FMG/NM interfaces. However, the first demonstration of robust SOT-induced magnetization switching in an FMG came only in 2017. Avci et al. have shown that the magnetization of an 8 nm-thick TmIG can be switched by SOTs generated by a 5 nm-thick Pt overlayer.[86] Through Hall effect measurements, the magnetization was found to switch between the up and down states after consecutive millisecond-long pulses of alternating polarity in the presence of an in-plane field (Fig. 5a-b). The switching polarity was found to change upon reversing the in-plane field, expectedly of the SOT-induced switching. Furthermore, the critical current density was characterized with a variety of in-plane fields and found to be inversely proportional to the external field strength (Fig. 5c-d) similar to previous reports.[38] The switching current densities through Pt were comparable to those used in metallic Pt/Co(~1nm) bilayers despite the large magnetic volume (8 nm) of TmIG. However, after considering the difference in the saturation magnetization -which is about one order of magnitude lower in FMGs with respect to Co- the switching efficiency in TmIG/Pt was found to be comparable to Pt/Co and other Pt-based metallic heterostructures.

Later on, the SOT-induced switching in TmIG/Pt was also demonstrated with a few nanosecond-long pulses and with a low in-plane field requirement of only 2 Oe in 2 μm-wide Hall bars[94] (Fig. 5e-f). Following these initial demonstrations, several other works have reported SOT-induced switching in TmIG/Pt, TbIG/Pt, YIG/Pt and Bi:YIG/Pt bilayers with similar or higher efficiencies.[57,95-98] In particular, Vélez et al. have reported local switching of magnetization in a continuous TmIG film by sending current through patterned Pt structures.[96] Local control of the magnetization in extended films is unique to magnetic insulators, and may find interest in the implementation and in-situ reconfiguration of synthetic magnetic structures for magnonic applications.

The SOT switching of TmIG is not limited to the use of Pt. Shao et al. have shown efficient switching in TmIG/W bilayers with TmIG thickness varying between 3.2 and 15 nm.[99] The switching polarity was found the opposite to that obtained in TmIG/Pt for a fixed current polarity, verifying the predominant SHE origin of the switching torques from Pt and W. Note that the SHE in Pt and W has an opposite sign. Switching in thick, bulk-like TmIG has demonstrated the significant potential of SOTs for controlling the magnetization in a wide thickness range. This study has also revealed that the SOT switching efficiency strongly depends on the TmIG thickness and that above 10 nm of TmIG, the SOT reaches the maximum efficiency (Fig. 6). It was argued that the spin current absorption is less efficient in thinner films due to the reduced magnetic moment density, which effectively reduces the exchange interaction between the current-induced accumulated spins and the local magnetic moments of TmIG near the interface, producing smaller torques. When the $M_s$ is close to the bulk limit, though, the SOT efficiency was maximized. Similar conclusions have been reached in another study by using different SHE metals grown on TmIG.[100] Interestingly, while switching of FMGs with PMA has been demonstrated in several different materials, SOT-induced switching of in-plane FMGs has remained little explored.[101] Limitations in suitable device fabrication with strong in-plane anisotropy and magnetization detection may be some of the reasons for the lack of such experiments. Overall, the various reports on FMG-based heterostructures have demonstrated that the SOTs act similarly upon magnetization in insulating magnets and hence revealing a significant potential for consideration in electrically-controlled magnetic devices.

## 5. Chiral Domain Walls and Their Current-Driven Motion in Ferrimagnetic Garnets

Magnetic DWs and their current-driven dynamics have been a prominent subject in magnetism and spintronics due to their intriguing physics and potential use in data storage and logic devices.[102,103] Earlier experiments of current-driven DW motion relied on less efficient STTs. Recently though, SOTs emerged as more efficient alternative to displace DWs, which, moreover, does not require the magnetic layer to be conducting.

SOT-induced DW motion is most effective in PMA layers possessing Néel type DWs due to the orthogonal alignment of DW internal spins and $\tau_{DL}$, which maximizes the effective field on the DW. Earlier, it was believed that in the standard PMA systems like Pt/Co, the DWs are of Bloch type governed by magnetostatics. Lately, though, it was understood that the interfacial Dzyaloshinskii-Moriya interaction (DMI) in Pt/Co and other similar NM/ferromagnet layers, gives rise to homochiral Néel DWs in structurally asymmetric multilayers.[104] Therefore, SOTs have been shown to move DWs in metallic heterostructures effectively.[21,22]

The efficient SOT-induced switching in FMGs, as well as the increased interest in the interfacial DMI, have triggered research into DWs and their current-induced dynamics in perpendicular FMGs. In 2019, Avci et al. have shown, through measurements of SOT-induced DW motion, that TmIG and TbIG possess DMI-stabilized left-handed Néel DWs.[105] A series of measurements with Pt and Cu/Pt capping layers on TmIG and TbIG indicated that the DMI predominantly originates from the FMG interface with the GGG substrate and not from the interface with the metal on top. This was the first direct demonstration of chiral magnetism occurring at a magnetic oxide interface. Shortly after, Vélez et al. and Ding et al. have reached the same conclusion that the GGG/TmIG substrate interface gives rise to the interfacial DMI and left-handed DWs in TmIG.[95,96] Vélez et al., moreover, have found, through nitrogen-vacancy center magnetometry measurements, that the Pt capping contributes to the DMI with an opposite sign; thereby concluded that the DMI contributions from the two interfaces compete with each other (see Fig. 7). Nevertheless, the reported DMI effective fields were only in the order of a few mT but strong enough to turn the equilibrium DW configuration from the Bloch-type to the partially or fully Néel-type.

Later on, Ding et al. and Caretta et al. have confirmed the interfacial origin of the DMI through TmIG thickness-dependent experiments.[100,106] Furthermore, Caretta et al. have reported that the DMI originates from the rare-earth orbital magnetism since the effect was absent in perpendicular Bi:YIG films but present in TmIG and TbIG grown on GGG substrates (Fig.8). Given the wide tunability of FMGs through composition and substrate choice, understanding the origins of DMI is an important step for designing materials with chiral spin textures. Despite experimental progress, though, the theoretical foundations of the interfacial DMI at oxide-oxide garnet interfaces remains yet to be established.

In the study mentioned above by Avci et al.,[105] it was also shown that the DMI-stabilized chiral DWs in TmIG can be moved in a DW track (Fig. 9a) by current injection through the Pt overlayers as fast as 800 m/s with current densities of the order of $1.2 \times 10^8$ A/cm$^2$ (Fig. 9b). The fast domain wall dynamics were a consequence of ferrimagnetic ordering and the absence of the precessional dynamics. In ferromagnetic systems such as Pt/Co, the DW velocity in the flow regime linearly increases as a function of current and saturates when the DL-SOT effective field becomes comparable to the DMI effective field.[104] This is due to the rotation of the DMI-stabilized DW spins towards the transverse direction (parallel to $\tau_{DL}$) at large excitations. However, in ferrimagnets, the antiferromagnetic coupling between the two sublattices prevents the net moment from precessing; hence the limiting factor is not anymore the DMI effective field but rather the much larger exchange field.[25] This intriguing physics makes the ferrimagnetic materials appealing for DW-based ultrafast spintronic devices.

In compensated FMGs, additional to the magnetic compensation temperature, there exists a second temperature at which the angular momentum is compensated and the magnetization dynamics resembles to that of antiferromagnets, allowing fast domain wall motion.[25,107] A recent study has shown that out-of-plane field assisted current-induced domain wall motion in GdIG with PMA can be as fast as 6000 m/s near the angular momentum compensation.[54] This result marks the highest domain wall velocity reported to date and demonstrates yet other degrees of freedom offered by FMGs for ultrafast spintronics.

Very recently, Caretta et al. have reported a fascinating feature of current-induced DW motion in FMGs[98]. They have shown that in Bi:YIG, DWs can be propelled at velocities in excess of 4300 m/s (Fig.9 c), the highest reported in any material so far without the need of an out-of-plane assisting field. The velocity was found to saturate to a universal limit above a critical current density and applied in-plane fields. Supported by analytical and atomistic modeling, this saturation anomaly was attributed to Lorentz contraction, a consequence of special relativity on the extremely fast-moving magnetic solitons. The reported velocities were calculated to be within 10% of the relativistic limit. These experiments open doors to realizing relativistic phenomena such as spin-wave Cherenkov radiation, higher-dimensional relativistic dynamics, and a host of related phenomena that previously seemed out of reach. And to achieve these behaviors in room-temperature, table-top experiments will hopefully lead to significant advances. Importantly, understanding the fundamental limits to the speeds of magnetic DWs and solitons, and the relation to the underlying material parameters can be used to maximize their speeds in future devices.

## 6. Signatures of Skyrmions in Ferrimagnetic Garnets

The exciting discovery of chiral magnetism at FMGs' interfaces and ultrafast DW motion have also accelerated the research into magnetic skyrmions in these materials. A skyrmion is a topologically protected magnetic texture stabilized by the DMI.[108] Skyrmions can be as small as a few nm in diameter up to hundreds of nm determined by the material parameters, and can be moved by SOTs,[27] Henceforth, they offer desirable prospects for efficient data storage, processing, and logic applications.[108]

So far, the direct observation of skyrmions in FMGs has been elusive. Though, strong evidence of skyrmions in TmIG and YIG has been found through the measurements of the topological Hall effect[109,110] (THE). The THE emerges due to the transverse deflection of electrons passing through the electromagnetic field generated by a skyrmion[111] (Fig. 10a). It gives a Hall signal contribution proportional to the skyrmion density in the (Hall cross area of the) film and is generally characterized by additional bumps overlapping the standard Hall resistance curves within a characteristic out-of-plane applied field range (Fig. 10b). Characterization of the THE as a function of FMG thickness, temperature and applied magnetic fields has revealed that there is a phase pocket in which the THE emerges (Fig. 10c). The common observation in such studies is that the THE occurs in ultrathin FMG films (<6 nm), particularly at higher temperatures, when the PMA becomes weaker. The collective understanding of these findings pinpoints the existence of skyrmions in ultrathin FMGs at the right combination of different material and experimental parameters.

Independently of the above studies, bubble domains of only a few hundred nm of diameter (similar to typical skyrmion size) have been observed in relatively thick (25.6 nm) TmIG by scanning transmission x-ray microscopy.[112] Photoemission electron microscopy analysis further showed that the DWs at the boundaries of the observed bubbles are of Bloch-type with arbitrary chirality. Such bubbles have been known since 1970s in micrometer-thick films,[113] but the current study marked their first nucleation, observation and detailed characterization in nanometer-thick FMGs. The established sample fabrication procedure and experimental scheme for observing such small magnetic textures, combined with the efforts of electrically characterizing skyrmions in ultrathin FMGs holds promise for the direct observation and engineering of insulating skyrmions in the near future.

## 7. Conclusions

The results summarized in this Review were obtained using only several FMGs and with a few years of experimental efforts. Yet, some impressive results such as highly competitive SOT-induced switching, interfacial chiral magnetism, ultrafast current-driven DW motion and relativistic current-induced dynamics have been achieved. The vast family of FMGs with many parameter tuning knobs combined with alternative SOT source materials creates a fertile ground for future studies and application prospects. Thus far, the SOT source has been limited to the SHE in Pt and W. There exist many other SOT materials and mechanisms including Weyl semimetals, topological insulators, two-dimensional electron gas interfaces, the interfacial

Rashba-Edelstein effect and more recently orbital Hall currents[114] are to be considered for more effective magnetic manipulation of FMGs in suitable devices.

The advances in the deposition of FMGs shall inevitably open the doors to investigations of magnetic coupling phenomena in multilayers. Thus far, only single FMG layers have been considered in the studies summarized here. Sequential deposition of FMGs with varying magnetic and structural properties may lead to novel properties and functionalities that may not be achievable by individual layers. For instance, a recent report showed antiferromagnetic coupling between YIG and GdIG.[115] Similarly coupled FMGs with PMA could reveal novel device concepts and architectures in electrically-controlled insulating spintronic devices.

The lack of in-depth understanding of the SOT-induced switching characteristics in FMGs opens new arenas for exploration. In all-metal systems, a significant effort has been spent to understand the SOT switching dynamics, device size and shape dependence, speed limits, etc. in the past couple of years, revealing outstanding results.[116-118] For instance, the state-of-the-art switching speed in all-metal systems are in the order of a few picoseconds,[119] some three orders of magnitude lower than that obtained in FMGs. Due to smaller Gilbert damping in FMGs in comparison to metallic ferromagnets, the switching dynamics in short timescales will be considerably different. FMGs are best suited to explore and understand the effect of the Gilbert damping on the switching, domain wall and skyrmion dynamics since it can be effectively tuned by the rare-earth choice. Efforts for understanding fast switching dynamics, device size and shape dependencies and the effect of the Gilbert damping in FMG-based devices are lagging behind and could reveal unpredictable twists.

Taken together, FMGs, particularly those possessing PMA, have emerged as highly promising materials for electrical manipulation of magnetization in spintronic and magnonic devices. We are only starting to become aware of the vast potential of FMGs for spintronic research and future applications. It is beyond doubt that FMG-based materials and associated physical phenomena will keep scientists busy for many years to come.

**Acknowledgements**

The author thanks Dr. Lucas Caretta and Martin Testa Anta for fruitful discussions. The author acknowledges funding from the European Research Council (ERC) under the European Union's Horizon 2020 research and innovation programme (project MAGNEPIC, grant agreement No. 949052).

Author email address: cavci@icmab.es

**References**

[1] F. Bertaut and F. Forrat, Acad. Sci. Paris **242**, 382 (1956).

[2] S. Geller and M. A. Gilleo, Acta Cryst. **10**, 239 (1957).

[3] G. Winkler, *Magnetic Garnets*. (Braunschweig, Vieweg, 1981), p.735.

[4] S. L. Blank and J. W. Nielsen, J. Cryst. Growth **17**, 302 (1972).

[5] J. L. Tomlinson and H. H. Wieder, Radio Electron. Eng. **45**, 725 (1975).

[6] D. Lacklison, G. Scott, A. Giles, J. Clarke, and R. Pearson, IEEE Trans. Magn. **13**, 973 (1977).

[7] B. Hill, IEEE Trans. Magn. **20**, 978 (1984).

[8] M. N. Baibich, J. M. Broto, A. Fert, F. Nguyen Van Dau, F. Petroff, P. Etienne, G. Creuzet, A. Friederich, and J. Chazelas, Phys. Rev. Lett. **61**, 2472 (1988).

[9] G. Binasch, P. Grunberg, F. Saurenbach, and W. Zinn, Phys. Rev. B **39** (7), 4828 (1989).

[10] P. F. Carcia, J. Appl. Phys. **63** (10), 5066 (1988).

[11] J. C. Slonczewski, J. Magn. Magn. Mater. **159** (1), L1 (1996).

[12] L. Berger, Phys. Rev. B **54** (13), 9353 (1996).

[13] Arne Brataas, Andrew D. Kent, and Hideo Ohno, Nat. Mater. **11** (5), 372 (2012).

[14] Claude Chappert, Albert Fert, and Frédéric Nguyen Van Dau, Nat. Mater. **6** (11), 813 (2007).

[15] A. A. Serga, A. V. Chumak, and B. Hillebrands, J. Phys. D: Appl. Phys. **43,** 264002 (2010).

[16] A. V. Chumak, V. I Vasyuchka, A. A Serga, and B. Hillebrands, Nat. Phys. **11**, 453 (2015).

[17] L. J. Cornelissen, J. Liu, R. A. Duine, J. Ben Youssef, and B. J. van Wees, Nat. Phys. **11**, 1022 (2015).

[18] Lucile Soumah, Nathan Beaulieu, Lilia Qassym, Cécile Carrétéro, Eric Jacquet, Richard Lebourgeois, Jamal Ben Youssef, Paolo Bortolotti, Vincent Cros, and Abdelmadjid Anane, Nat. Commun. **9** , 3355 (2018).

[19] Chuanpu Liu, Jilei Chen, Tao Liu, Florian Heimbach, Haiming Yu, Yang Xiao, Junfeng Hu, Mengchao Liu, Houchen Chang, Tobias Stueckler, Sa Tu, Youguang Zhang, Yan Zhang, Peng Gao, Zhimin Liao, Dapeng Yu, Ke Xia, Na Lei, Weisheng Zhao, and Mingzhong Wu, Nat. Commun. **9**, 738 (2018).

[20] P. Xu, K. Xia, C. Gu, L. Tang, H. Yang, and J. Li, Nat. Nanotech. **3**, 97 (2008).

[21] K. S. Ryu, L. Thomas, S. H. Yang, and S. Parkin, Nat. Nanotech. **8**, 527 (2013).

[22] S. Emori, U. Bauer, S. M. Ahn, E. Martinez, and G. S. Beach, Nat. Mater. **12**, 611 (2013).

[23] G. S. D. Beach, M. Tsoi, and J. L. Erskine, J. Magn. Magn. Mater. **320**, 1272 (2008).

[24] P. P. Haazen, E. Mure, J. H. Franken, R. Lavrijsen, H. J. Swagten, and B. Koopmans, Nat. Mater. **12**, 299 (2013).

[25] L. Caretta, M. Mann, F. Buttner, K. Ueda, B. Pfau, C. M. Gunther, P. Hessing, A. Churikova, C. Klose, M. Schneider, D. Engel, C. Marcus, D. Bono, K. Bagschik, S. Eisebitt, and G. S. D. Beach, Nat. Nanotech. **13**, 1154 (2018).

[26] I. M. Miron, T. Moore, H. Szambolics, L. D. Buda-Prejbeanu, S. Auffret, B. Rodmacq, S. Pizzini, J. Vogel, M. Bonfim, A. Schuhl, and G. Gaudin, Nat. Mater. **10**, 419 (2011).

[27] S. Woo, K. Litzius, B. Kruger, M. Y. Im, L. Caretta, K. Richter, M. Mann, A. Krone, R. M. Reeve, M. Weigand, P. Agrawal, I. Lemesh, M. A. Mawass, P. Fischer, M. Klaui, and G. S. Beach, Nat. Mater. **15**, 501 (2016).

[28] A. Fert, V. Cros, and J. Sampaio, Nat Nanotechnol **8** (3), 152 (2013).

[29] Ümit Özgür, Yahya Alivov, and Hadis Morkoç, J. Mater. Sci.: Mater. Electron. **20**, 789 (2009).

[30] J. J. Bauer, E. R. Rosenberg, and C. A. Ross, Appl. Phys. Lett. **114**, 052403 (2019).

[31] Y. Kajiwara, K. Harii, S. Takahashi, J. Ohe, K. Uchida, M. Mizuguchi, H. Umezawa, H. Kawai, K. Ando, K. Takanashi, S. Maekawa, and E. Saitoh, Nature **464**, 262 (2010).

[32] B. Heinrich, C. Burrowes, E. Montoya, B. Kardasz, E. Girt, Y. Y. Song, Y. Sun, and M. Wu, Phys. Rev. Lett. **107**, 066604 (2011).

[33] Ken-ichi Uchida, Hiroto Adachi, Takeru Ota, Hiroyasu Nakayama, Sadamichi Maekawa, and Eiji Saitoh, Appl. Phys. Lett. **97**, 172505 (2010).

[34] H. Nakayama, M. Althammer, Y. T. Chen, K. Uchida, Y. Kajiwara, D. Kikuchi, T. Ohtani, S. Geprags, M. Opel, S. Takahashi, R. Gross, G. E. Bauer, S. T. Goennenwein, and E. Saitoh, Phys. Rev. Lett. **110**, 206601 (2013).

[35] A. Manchon, J. Železný, I. M Miron, T. Jungwirth, J. Sinova, A. Thiaville, K. Garello, and P. Gambardella, Rev. Mod. Phys. **91**, 035004 (2019).

[36] Ioan Mihai Miron, Kevin Garello, Gilles Gaudin, Pierre-Jean Zermatten, Marius V. Costache, Stéphane Auffret, Sébastien Bandiera, Bernard Rodmacq, Alain Schuhl, and Pietro Gambardella, Nature **476**, 189 (2011).

[37] Can Onur Avci, Kevin Garello, Ioan Mihai Miron, Gilles Gaudin, Stéphane Auffret, Olivier Boulle, and Pietro Gambardella, Appl. Phys. Lett. **100**, 212404 (2012).

[38] Can Onur Avci, Kevin Garello, Corneliu Nistor, Sylvie Godey, Belén Ballesteros, Aitor Mugarza, Alessandro Barla, Manuel Valvidares, Eric Pellegrin, Abhijit Ghosh, Ioan Mihai

Miron, Olivier Boulle, Stephane Auffret, Gilles Gaudin, and Pietro Gambardella, Phys. Rev. B **89**, 214419 (2014).

[39] Luqiao Liu, O. J. Lee, T. J. Gudmundsen, D. C. Ralph, and R. A. Buhrman, Phys. Rev. Lett. **109**, 096602 (2012).

[40] Luqiao Liu, Chi-Feng Pai, Y. Li, H. W. Tseng, D. C. Ralph, and R. A. Buhrman, Science **336**, 555 (2012).

[41] V. E. Demidov, S. Urazhdin, A. Anane, V. Cros, and S. O. Demokritov, J. Appl. Phys. **127**, 170901 (2020).

[42] G. P. Rodrigue, IEEE Proc. **76**, 121 (1988).

[43] Yiyan Sun, Young-Yeal Song, Houchen Chang, Michael Kabatek, Michael Jantz, William Schneider, Mingzhong Wu, Helmut Schultheiss, and Axel Hoffmann, Appl. Phys. Lett. **101**, 152405 (2012).

[44] Tao Liu, Houchen Chang, Vincent Vlaminck, Yiyan Sun, Michael Kabatek, Axel Hoffmann, Longjiang Deng, and Mingzhong Wu, J. Appl. Phys. **115**, 17A501 (2014).

[45] M. C. Onbasli, A. Kehlberger, D. H. Kim, G. Jakob, M. Kläui, A. V. Chumak, B. Hillebrands, and C. A. Ross, APL Mater. **2**, 106102 (2014).

[46] C. N. Wu, C. C. Tseng, K. Y. Lin, C. K. Cheng, S. L. Yeh, Y. T. Fanchiang, M. Hong, and J. Kwo, AIP Adv. **8**, 055904 (2017).

[47] Gilvânia Vilela, Hang Chi, Gregory Stephen, Charles Settens, Preston Zhou, Yunbo Ou, Dhavala Suri, Don Heiman, and Jagadeesh S. Moodera, J. Appl. Phys. **127**, 115302 (2020).

[48] M. Kubota, K. Shibuya, Y. Tokunaga, F. Kagawa, A. Tsukazaki, Y. Tokura, and M. Kawasaki, J. Magn. Magn. Mater. **339**, 63 (2013).

[49] Masashi Kubota, Atsushi Tsukazaki, Fumitaka Kagawa, Keisuke Shibuya, Yusuke Tokunaga, Masashi Kawasaki, and Yoshinori Tokura, Appl. Phys. Exp. **5**, 103002 (2012).

[50] Andy Quindeau, Can O. Avci, Wenqing Liu, Congli Sun, Maxwell Mann, Astera S. Tang, Mehmet C. Onbasli, David Bono, Paul M. Voyles, Yongbing Xu, Jason Robinson, Geoffrey S. D. Beach, and Caroline A. Ross, Adv. Electron. Mater. **3**, 1600376 (2017).

[51] Oana Ciubotariu, Anna Semisalova, Kilian Lenz, and Manfred Albrecht, Sci. Rep. **9**, 17474 (2019).

[52] Victor H. Ortiz, Mohammed Aldosary, Junxue Li, Yadong Xu, Mark I. Lohmann, Pathikumar Sellappan, Yasuhiro Kodera, Javier E. Garay, and Jing Shi, APL Mater. **6**, 121113 (2018).

[53] Ethan R. Rosenberg, Lukáš Beran, Can O. Avci, Cyrus Zeledon, Bingqian Song, Claudio Gonzalez-Fuentes, Johannes Mendil, Pietro Gambardella, Martin Veis, Carlos Garcia, Geoffrey S. D. Beach, and Caroline A. Ross, Phys. Rev. Mater. **2**, 094405 (2018).

54 H.A. Zhou, Y. Dong, T. Xu, K. Xu, L. Sánchez-Tejerina, L. Zhao, Y Ba, P. Gargiani, M. Valvidares, Y. Zhao, M. Carpentieri, O. A. Tretiakov, X. Zhong, G. Finocchio, S. W. Kim, and W. Jiang, arXiv:1912.01775.

55 Jianbo Fu, Muxin Hua, Xin Wen, Mingzhu Xue, Shilei Ding, Meng Wang, Pu Yu, Shunquan Liu, Jingzhi Han, Changsheng Wang, Honglin Du, Yingchang Yang, and Jinbo Yang, Appl. Phys. Lett. **110**, 202403 (2017).

56 Jackson J. Bauer, Ethan R. Rosenberg, Subhajit Kundu, K. Andre Mkhoyan, Patrick Quarterman, Alexander J. Grutter, Brian J. Kirby, Julie A. Borchers, and Caroline A. Ross, Adv. Electron. Mater. **6**, 1900820 (2020).

57 Jinjun Ding, Chuanpu Liu, Yuejie Zhang, Uppalaiah Erugu, Zhiyong Quan, Rui Yu, Ethan McCollum, Songyu Mo, Sheng Yang, Haifeng Ding, Xiaohong Xu, Jinke Tang, Xiaofei Yang, and Mingzhong Wu, Phys. Rev. Appl. **14**, 014017 (2020).

58 Aidan J. Lee, Adam S. Ahmed, Brendan A McCullian, Side Guo, Menglin Zhu, Sisheng Yu, Patrick M Woodward, Jinwoo Hwang, P. Chris Hammel, and Fengyuan Yang, Phys. Rev. Lett. **124**, 257202 (2020).

59 Alexandr Chernyshov, Mason Overby, Xinyu Liu, Jacek K. Furdyna, Yuli Lyanda-Geller, and Leonid P. Rokhinson, Nat. Phys. **5**, 656 (2009).

60 I. M. Miron, G. Gaudin, S. Auffret, B. Rodmacq, A. Schuhl, S. Pizzini, J. Vogel, and P. Gambardella, Nat. Mater. **9**, 230 (2010).

61 K. Garello, I. M. Miron, C. O. Avci, F. Freimuth, Y. Mokrousov, S. Blugel, S. Auffret, O. Boulle, G. Gaudin, and P. Gambardella, Nat. Nanotech. **8**, 587 (2013).

62 J. Kim, J. Sinha, M. Hayashi, M. Yamanouchi, S. Fukami, T. Suzuki, S. Mitani, and H. Ohno, Nat. Mater. **12**, 240 (2013).

63 Chi-Feng Pai, Luqiao Liu, Y. Li, H. W. Tseng, D. C. Ralph, and R. A. Buhrman, Appl. Phys. Lett. **101**, 122404 (2012).

64 Jiahao Han, A. Richardella, Saima A. Siddiqui, Joseph Finley, N. Samarth, and Luqiao Liu, Phys. Rev. Lett. **119**, 077702 (2017).

65 Peng Li, James Kally, Steven S. L. Zhang, Timothy Pillsbury, Jinjun Ding, Gyorgy Csaba, Junjia Ding, J. S. Jiang, Yunzhi Liu, Robert Sinclair, Chong Bi, August DeMann, Gaurab Rimal, Wei Zhang, Stuart B. Field, Jinke Tang, Weigang Wang, Olle G. Heinonen, Valentine Novosad, Axel Hoffmann, Nitin Samarth, and Mingzhong Wu, Sci. Adv. **5**, eaaw3415 (2019).

66 Qiming Shao, Guoqiang Yu, Yann-Wen Lan, Yumeng Shi, Ming-Yang Li, Cheng Zheng, Xiaodan Zhu, Lain-Jong Li, Pedram Khalili Amiri, and Kang L. Wang, Nano Lett. **16**, 7514 (2016).

67 Shuyuan Shi, Shiheng Liang, Zhifeng Zhu, Kaiming Cai, Shawn D. Pollard, Yi Wang, Junyong Wang, Qisheng Wang, Pan He, Jiawei Yu, Goki Eda, Gengchiau Liang, and Hyunsoo Yang, Nat. Nanotech. **14**, 945 (2019).

[68] Hongyu An, Takeo Ohno, Yusuke Kanno, Yuito Kageyama, Yasuaki Monnai, Hideyuki Maki, Ji Shi, and Kazuya Ando, Sci. Adv. **4**, eaar2250 (2018).

[69] Kai-Uwe Demasius, Timothy Phung, Weifeng Zhang, Brian P. Hughes, See-Hun Yang, Andrew Kellock, Wei Han, Aakash Pushp, and Stuart S. P. Parkin, Nat. Commun. **7**, 10644 (2016).

[70] Junxiao Feng, Eva Grimaldi, Can Onur Avci, Manuel Baumgartner, Giovanni Cossu, Antonella Rossi, and Pietro Gambardella, Phys. Rev. Appl. **13**, 044029 (2020).

[71] Lijun Zhu, Daniel C. Ralph, and Robert A. Buhrman, Phys. Rev. Appl. **10**, 031001 (2018).

[72] Lijun Zhu, Kemal Sobotkiewich, Xin Ma, Xiaoqin Li, Daniel C. Ralph, and Robert A. Buhrman, Adv. Func. Mater. **29**, 1805822 (2019).

[73] A. R. Mellnik, J. S. Lee, A. Richardella, J. L. Grab, P. J. Mintun, M. H. Fischer, A. Vaezi, A. Manchon, E. A. Kim, N. Samarth, and D. C. Ralph, Nature **511**, 449 (2014).

[74] D. MacNeill, G. M. Stiehl, M. H. D. Guimaraes, R. A. Buhrman, J. Park, and D. C. Ralph, Nat. Phys. **13**, 300 (2017).

[75] Satoshi Iihama, Tomohiro Taniguchi, Kay Yakushiji, Akio Fukushima, Yoichi Shiota, Sumito Tsunegi, Ryo Hiramatsu, Shinji Yuasa, Yoshishige Suzuki, and Hitoshi Kubota, Nat. Electron. **1**, 120 (2018).

[76] Christopher Safranski, Eric A. Montoya, and Ilya N. Krivorotov, Nat. Nanotech. **14**, 27 (2019).

[77] S. C. Baek, V. P. Amin, Y. W. Oh, G. Go, S. J. Lee, G. H. Lee, K. J. Kim, M. D. Stiles, B. G. Park, and K. J. Lee, Nat. Mater. **17**, 509 (2018).

[78] Minh-Hai Nguyen and Chi-Feng Pai, APL Mater. **9**, 030902 (2021).

[79] Masamitsu Hayashi, Junyeon Kim, Michihiko Yamanouchi, and Hideo Ohno, Phys. Rev. B **89**, 144425 (2014).

[80] Can Onur Avci, Kevin Garello, Mihai Gabureac, Abhijit Ghosh, Andreas Fuhrer, Santos F. Alvarado, and Pietro Gambardella, Phys. Rev. B **90,** 224427 (2014).

[81] Abhijit Ghosh, Kevin Garello, Can Onur Avci, Mihai Gabureac, and Pietro Gambardella, Phys. Rev. Appl. **7**, 014004 (2017).

[82] Matthias Althammer, Sibylle Meyer, Hiroyasu Nakayama, Michael Schreier, Stephan Altmannshofer, Mathias Weiler, Hans Huebl, Stephan Geprägs, Matthias Opel, Rudolf Gross, Daniel Meier, Christoph Klewe, Timo Kuschel, Jan-Michael Schmalhorst, Günter Reiss, Liming Shen, Arunava Gupta, Yan-Ting Chen, Gerrit E. W. Bauer, Eiji Saitoh, and Sebastian T. B. Goennenwein, Phys. Rev. B **87**, 224401 (2013).

[83] Miren Isasa, Amilcar Bedoya-Pinto, Saül Vélez, Federico Golmar, Florencio Sánchez, Luis E. Hueso, Josep Fontcuberta, and Fèlix Casanova, Appl. Phys. Lett. **105**, 142402 (2014).

[84] N. Vlietstra, J. Shan, V. Castel, B. J. van Wees, and J. Ben Youssef, Phys. Rev. B **87**, 184421 (2013).

[85] Yan-Ting Chen, Saburo Takahashi, Hiroyasu Nakayama, Matthias Althammer, Sebastian T. B. Goennenwein, Eiji Saitoh, and Gerrit E. W. Bauer, Phys. Rev. B **87**, 144411 (2013).

[86] C. O. Avci, A. Quindeau, C. F. Pai, M. Mann, L. Caretta, A. S. Tang, M. C. Onbasli, C. A. Ross, and G. S. Beach, Nat. Mater. **16**, 309 (2017).

[87] Can Onur Avci, Andy Quindeau, Maxwell Mann, Chi-Feng Pai, Caroline A. Ross, and Geoffrey S. D. Beach, Phys. Rev. B **95**, 115428 (2017).

[88] Junxue Li, Guoqiang Yu, Chi Tang, Yizhou Liu, Zhong Shi, Yawen Liu, Aryan Navabi, Mohammed Aldosary, Qiming Shao, Kang L. Wang, Roger Lake, and Jing Shi, Phys. Rev. B **95**, 241305(R) (2017).

[89] W. X. Wang, S. H. Wang, L. K. Zou, J. W. Cai, Z. G. Sun, and J. R. Sun, Appl. Phys. Lett. **105**, 182403 (2014).

[90] N. Vlietstra, J. Shan, B. J. van Wees, M. Isasa, F. Casanova, and J. Ben Youssef, Phys. Rev. B **90**, 174436 (2014).

[91] Joseph Sklenar, Wei Zhang, Matthias B. Jungfleisch, Wanjun Jiang, Houchen Chang, John E. Pearson, Mingzhong Wu, John B. Ketterson, and Axel Hoffmann, Phys. Rev. B **92**, 174406 (2015).

[92] Takahiro Chiba, Michael Schreier, Gerrit E. W. Bauer, and Saburo Takahashi, J. Appl. Phys. **117**, 17C715 (2015).

[93] Takahiro Chiba, Gerrit E. W Bauer, and Saburo Takahashi, Phys. Rev. Appl. **2**, 034003 (2014).

[94] Can Onur Avci, Ethan Rosenberg, Manuel Baumgartner, Lukáš Beran, Andy Quindeau, Pietro Gambardella, Caroline A. Ross, and Geoffrey S. D. Beach, Appl. Phys. Lett. **111**, 072406 (2017).

[95] Shilei Ding, Andrew Ross, Romain Lebrun, Sven Becker, Kyujoon Lee, Isabella Boventer, Souvik Das, Yuichiro Kurokawa, Shruti Gupta, Jinbo Yang, Gerhard Jakob, and Mathias Kläui, Phys. Rev. B **100**, 100406 (2019).

[96] Saül Vélez, Jakob Schaab, Martin S. Wörnle, Marvin Müller, Elzbieta Gradauskaite, Pol Welter, Cameron Gutgsell, Corneliu Nistor, Christian L. Degen, Morgan Trassin, Manfred Fiebig, and Pietro Gambardella, Nat. Commun. **10**, 4750 (2019).

[97] Huanjian Chen, Dashuai Cheng, Huanglin Yang, Daike Wang, Shiming Zhou, Zhong Shi, and Xuepeng Qiu, Appl. Phys. Lett. **116**, 112401 (2020).

[98] Lucas Caretta, Se-Hyeok Oh, Takian Fakhrul, Dong-Kyu Lee, Byung Hun Lee, Se Kwon Kim, Caroline A. Ross, Kyung-Jin Lee, and Geoffrey S. D. Beach, Science **370**, 1438 (2020).

[99] Qiming Shao, Chi Tang, Guoqiang Yu, Aryan Navabi, Hao Wu, Congli He, Junxue Li, Pramey Upadhyaya, Peng Zhang, Seyed Armin Razavi, Qing Lin He, Yawen Liu, Pei Yang, Se Kwon Kim, Cheng Zheng, Yizhou Liu, Lei Pan, Roger K. Lake, Xiufeng Han, Yaroslav Tserkovnyak, Jing Shi, and Kang L. Wang, Nat. Commun. **9**, 3612 (2018).

[100] Shilei Ding, Lorenzo Baldrati, Andrew Ross, Zengyao Ren, Rui Wu, Sven Becker, Jinbo Yang, Gerhard Jakob, Arne Brataas, and Mathias Kläui, Phys. Rev. B **102**, 054425 (2020).

[101] Johannes Mendil, Morgan Trassin, Qingqing Bu, Manfred Fiebig, and Pietro Gambardella, Appl. Phys. Lett. **114**, 172404 (2019).

[102] Stuart S. P. Parkin, Masamitsu Hayashi, and Luc Thomas, Science **320**, 190 (2008).

[103] Z. Luo, A. Hrabec, T. P. Dao, G. Sala, S. Finizio, J. Feng, S. Mayr, J. Raabe, P. Gambardella, and L. J. Heyderman, Nature **579**, 214 (2020).

[104] André Thiaville, Stanislas Rohart, Émilie Jué, Vincent Cros, and Albert Fert, Europhys. Lett. **100**, 57002 (2012).

[105] C. O. Avci, E. Rosenberg, L. Caretta, F. Buttner, M. Mann, C. Marcus, D. Bono, C. A. Ross, and G. S. D. Beach, Nat. Nanotech. **14**, 561 (2019).

[106] Lucas Caretta, Ethan Rosenberg, Felix Büttner, Takian Fakhrul, Pierluigi Gargiani, Manuel Valvidares, Zhen Chen, Pooja Reddy, David A. Muller, Caroline A. Ross, and Geoffrey S. D. Beach, Nat. Commun. **11**, 1090 (2020).

[107] Kab-Jin Kim, Se Kwon Kim, Yuushou Hirata, Se-Hyeok Oh, Takayuki Tono, Duck-Ho Kim, Takaya Okuno, Woo Seung Ham, Sanghoon Kim, Gyoungchoon Go, Yaroslav Tserkovnyak, Arata Tsukamoto, Takahiro Moriyama, Kyung-Jin Lee, and Teruo Ono, Nat. Mater. **16**, 1187 (2017).

[108] Albert Fert, Nicolas Reyren, and Vincent Cros, Nat. Rev. Mater. **2**, 17031 (2017).

[109] Qiming Shao, Yawen Liu, Guoqiang Yu, Se Kwon Kim, Xiaoyu Che, Chi Tang, Qing Lin He, Yaroslav Tserkovnyak, Jing Shi, and Kang L. Wang, Nat.Electron. **2**, 182 (2019).

[110] Adam S. Ahmed, Aidan J. Lee, Nuria Bagués, Brendan A. McCullian, Ahmed M. A. Thabt, Avery Perrine, Po-Kuan Wu, James R. Rowland, Mohit Randeria, P. Chris Hammel, David W. McComb, and Fengyuan Yang, Nano Lett. **19**, 5683 (2019).

[111] Steven S. L. Zhang and Olle Heinonen, Phys. Rev. B **97**, 134401 (2018).

[112] Felix Büttner, Mohamad A. Mawass, Jackson Bauer, Ethan Rosenberg, Lucas Caretta, Can Onur Avci, Joachim Gräfe, Simone Finizio, C. A. F. Vaz, Nina Novakovic, Markus Weigand, Kai Litzius, Johannes Förster, Nick Träger, Felix Groß, Daniel Suzuki, Mantao Huang, Jason Bartell, Florian Kronast, Jörg Raabe, Gisela Schütz, Caroline A. Ross, and Geoffrey S. D. Beach, Phys. Rev. Mater. **4**, 011401 (2020).

[113] D. M. Heinz, P. J. Besser, J. M. Owens, J. E. Mee, and G. R. Pulliam, J. Appl. Phys. **42**, 1243 (1971).

[114] Shilei Ding, Andrew Ross, Dongwook Go, Lorenzo Baldrati, Zengyao Ren, Frank Freimuth, Sven Becker, Fabian Kammerbauer, Jinbo Yang, Gerhard Jakob, Yuriy Mokrousov, and Mathias Kläui, Phys. Rev. Lett. **125**, 177201 (2020).

[115] Juan M. Gomez-Perez, Saül Vélez, Lauren McKenzie-Sell, Mario Amado, Javier Herrero-Martín, Josu López-López, S. Blanco-Canosa, Luis E. Hueso, Andrey Chuvilin, Jason W. A. Robinson, and Fèlix Casanova, Phys. Rev. Appl. **10**, 044046 (2018).

[116] M. Baumgartner, K. Garello, J. Mendil, C. O. Avci, E. Grimaldi, C. Murer, J. Feng, M. Gabureac, C. Stamm, Y. Acremann, S. Finizio, S. Wintz, J. Raabe, and P. Gambardella, Nat. Nanotech. **12**, 980 (2017).

[117] Kevin Garello, Can Onur Avci, Ioan Mihai Miron, Manuel Baumgartner, Abhijit Ghosh, Stéphane Auffret, Olivier Boulle, Gilles Gaudin, and Pietro Gambardella, Appl. Phys. Lett. **105**, 212402 (2014).

[118] A. Kurenkov, C. Zhang, S. DuttaGupta, S. Fukami, and H. Ohno, Appl. Phys. Lett. **110**, 092410 (2017).

[119] Kaushalya Jhuria, Julius Hohlfeld, Akshay Pattabi, Elodie Martin, Aldo Ygnacio Arriola Córdova, Xinping Shi, Roberto Lo Conte, Sebastien Petit-Watelot, Juan Carlos Rojas-Sanchez, Gregory Malinowski, Stéphane Mangin, Aristide Lemaître, Michel Hehn, Jeffrey Bokor, Richard B. Wilson, and Jon Gorchon, Nat. Electron. **3**, 680 (2020).

**Figures**

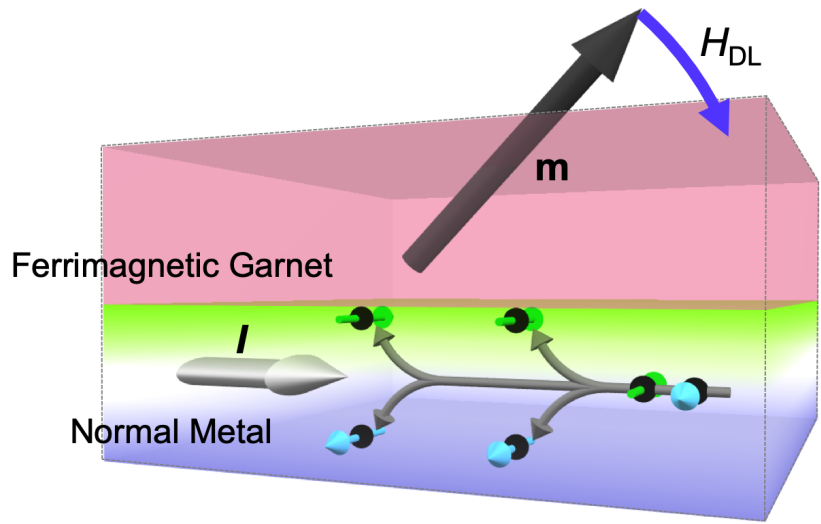

**Figure 1** – (Color online) Switching mechanism of the perpendicular magnetization in an FMG by using the SHE. A charge current through the normal metal (e.g., Pt) can generate a spin accumulation at the interface with the FMG, which then generates a damping-like SOT on the magnetization. The resulting effective field ($H_{DL}$) can switch the magnetization between the up and down state in the presence of an in-plane field applied along the current injection direction that breaks the rotational symmetry of $H_{DL}$. (Adapted from Ref. 87)

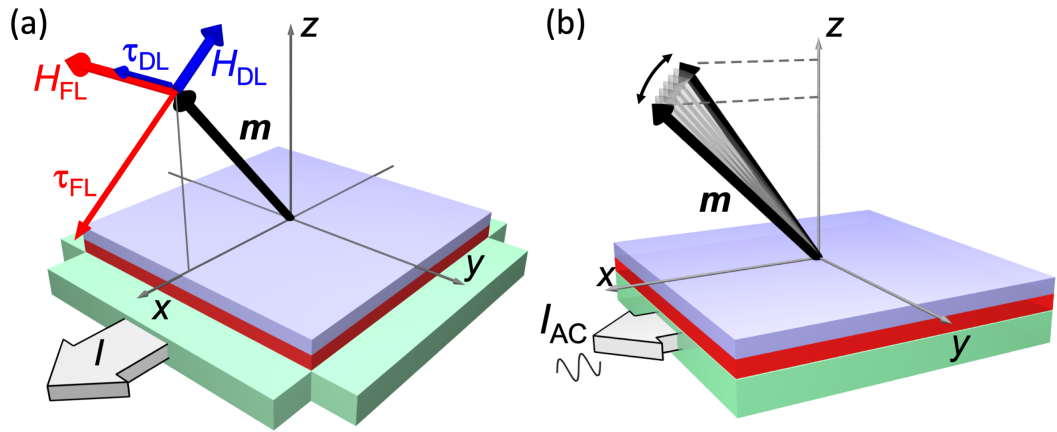

**Figure 2** – (Color online) **(a)** Damping-like and field-like SOTs ($\tau_{DL}$, $\tau_{FL}$), and their associated effective fields ($H_{DL}$, $H_{FL}$) for the given magnetization and current injection geometry. Note that the torque/field signs and amplitudes are arbitrary as they are material-dependent. **(b)** A visualization of the harmonic measurements. The magnetization vector oscillates with an a.c. current injection due to the oscillating SOTs and produce harmonic Hall voltage signals. (Adapted from Ref. 38)

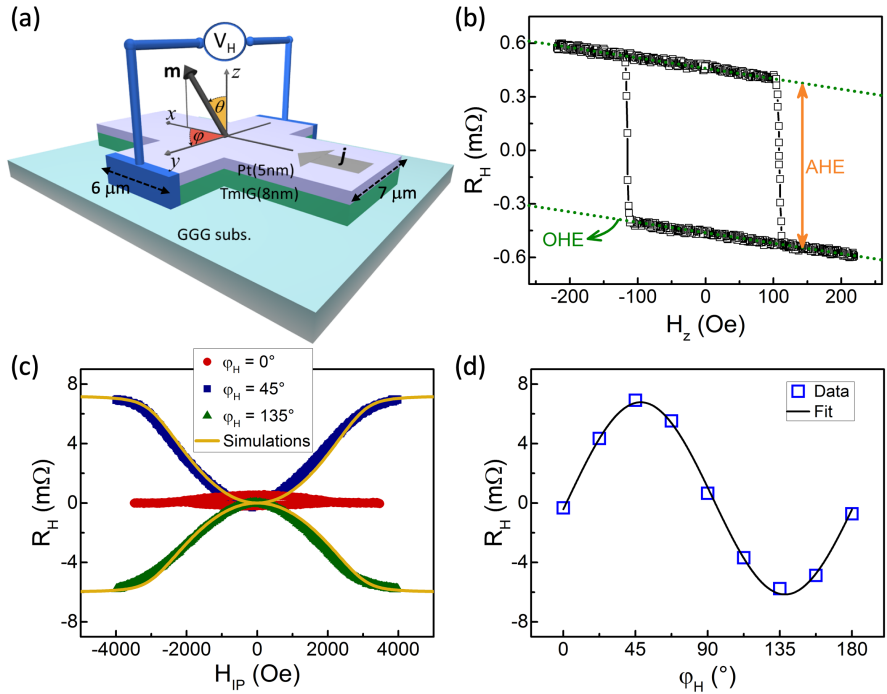

**Figure 3** – (Color online) **(a)** Schematic of a setup for the Hall effect measurement on a device made of TmIG/Pt. **(b)** Typical Hall resistance signal for TmIG/Pt layers with PMA, recorded during an out-of-plane field sweep, consisting of the anomalous and ordinary Hall effect contributions as shown. **(c)** Hall resistance measured with an in-plane field ($H_{IP}$) sweep applied at various angles. The curves at 45° and 135° represent the typical SMR behavior where a large positive and negative signal appears, and **m** is fully rotated in-plane. Orange curves are macrospin simulations based on experimental parameters and estimated PMA of about 2700 Oe. **(d)** Hall resistance signal amplitude measured at various in-plane angles reveals the expected $\sin 2\varphi$ behavior (see Eq. 1). (Adapted from Ref. 86)

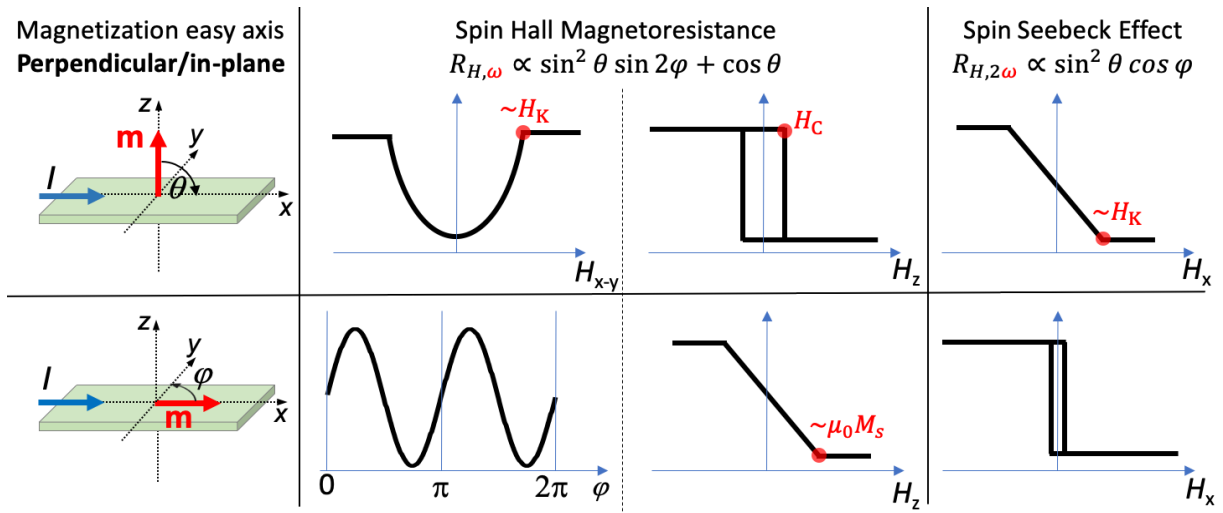

**Figure 4** – An electrical detection toolbox for magnetic insulators having out-plane (top row) or in-plane (bottom row) equilibrium magnetization orientation.

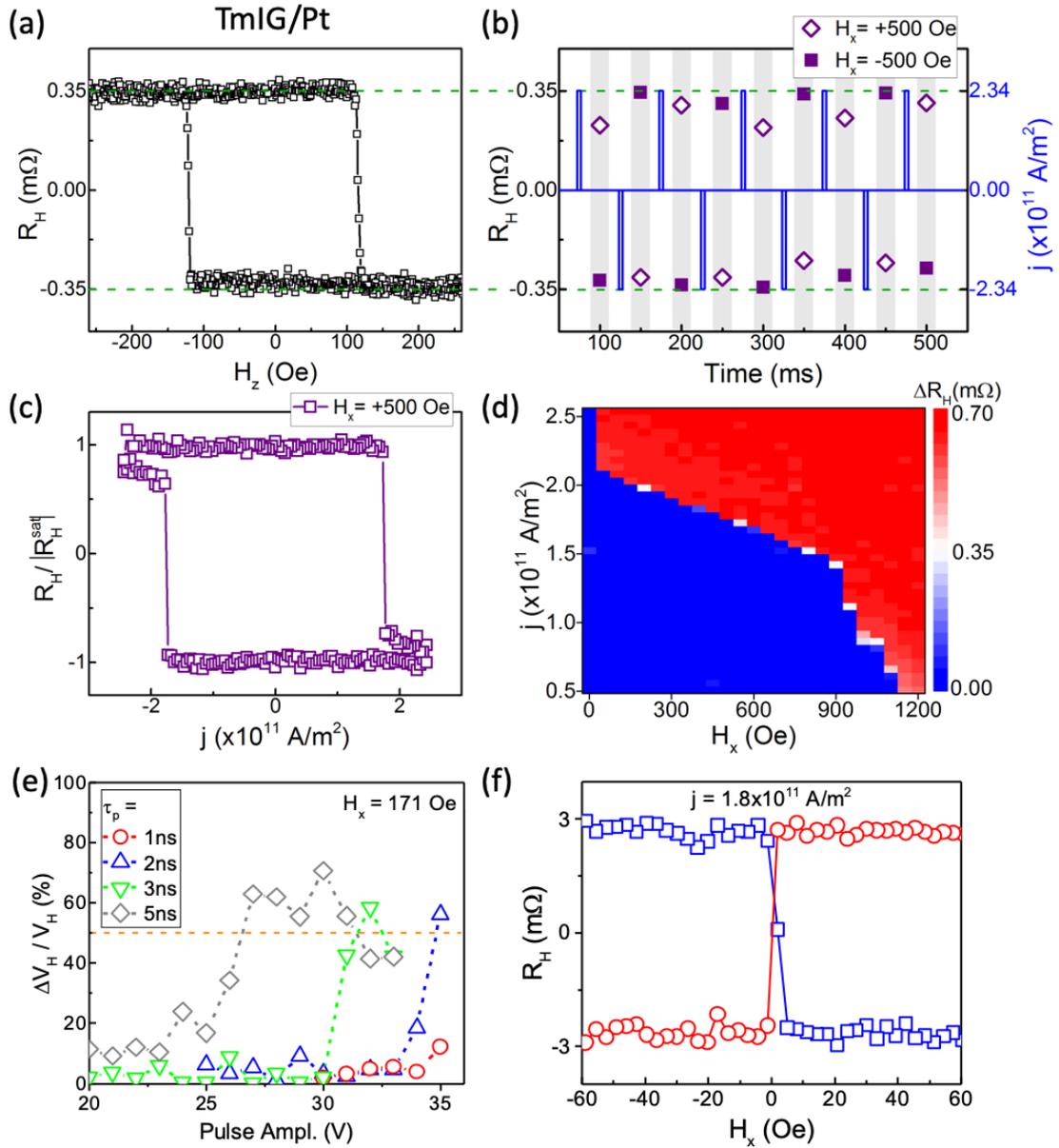

**Figure 5** – (Color online) Current-induced magnetization switching experiments in TmIG/Pt bilayers on Hall bar devices similar to the one in Fig. 3a. **(a)** Reference Hall resistance signal after subtraction of the OHE slope. **(b)** Magnetization switching experiments by consecutive application of 5 ms-long pulses through the device in the presence of an in-plane field of +/- 500 Oe. The switching polarity reverses upon inverting the field polarity, as expected of the SOT-induced switching behavior. **(c)** A current sweep in the presence of a constant in-plane field, showing reversible switching at critical current density of about $1.8 \times 10^{11}$ A/m$^2$. **(d)** Switching phase diagram constructed based on measurements reported in (c) with different in-plane fields. **(e)** Partial switching of TmIG by electrical pulses of 1-5 ns length and 25-35 V

amplitude. 100% corresponds to the full switching of the Hall cross area and 0% corresponds to no switching. **(f)** In-plane field dependence of the switching characterized by stepping the field and applying positive and negative pulses and measuring $R_H$ consecutively. These measurements have revealed an in-plane field requirement as low as 2 Oe for switching TmIG. (Adapted from Refs. 86 and 94)

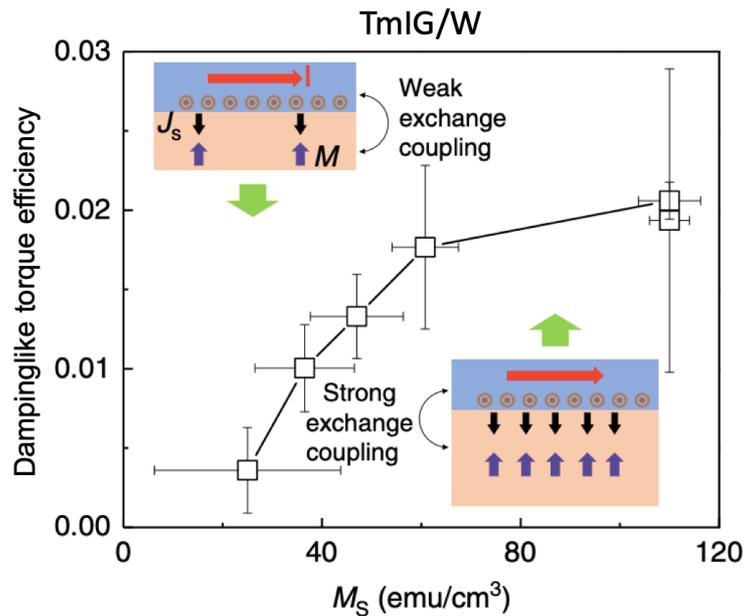

**Figure 6** – (Color online) TmIG saturation magnetization ($M_s$) dependence of the damping-like SOT efficiency in TmIG/W bilayers. The enhanced torque efficiency at larger $M_s$ (which is obtained at larger TmIG thicknesses) is attributed to the stronger exchange coupling between the SHE-induced interface spin accumulation and the local magnetization in TmIG. (Adapted from Ref. 99)

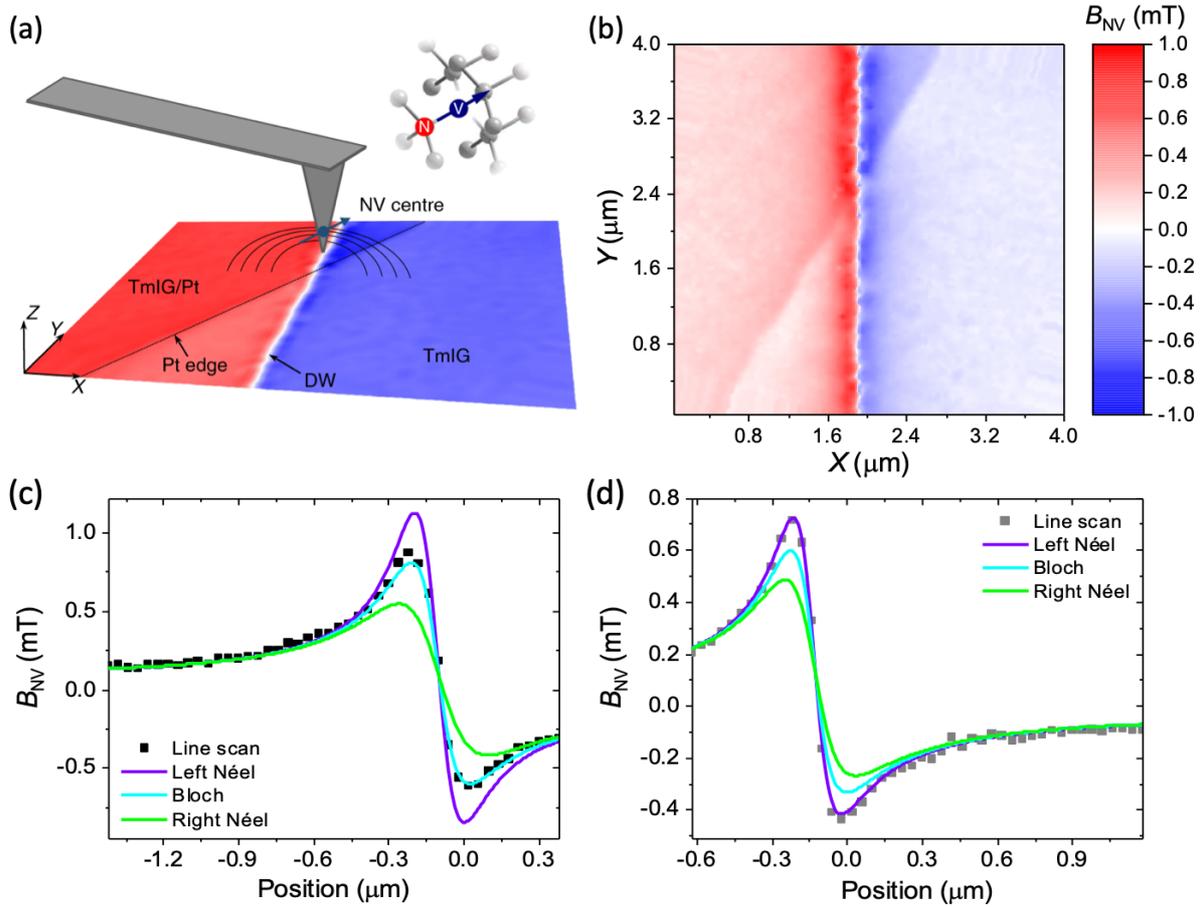

**Figure 7** – (Color online) Nitrogen-vacancy center microscopy characterization of domain wall internal magnetization in TmIG and TmIG/Pt bilayers. **(a)** Schematics of the experimental setup and the DW types. **(b)** Color map of the magnetization orientation measured by magnetic stray fields ($B_{NV}$) showing a straight vertical domain wall crossing the Pt edge. **(c,d)** $B_{NV}$ measured along the line scans on the TmIG/Pt and bare TmIG regions. Fits showing that the domain wall in TmIG/Pt is a Bloch-type while in TmIG only it is a left-handed Néel type. (Adapted from Ref. 96)

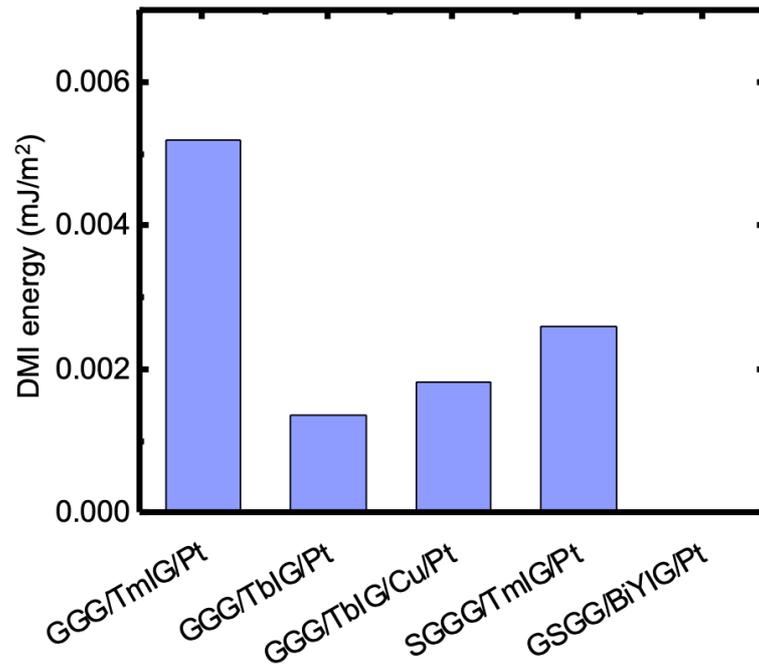

**Figure 8** – (Color online) The interfacial DMI energy measured in various FMG-based systems. (Adapted from Ref. 106)

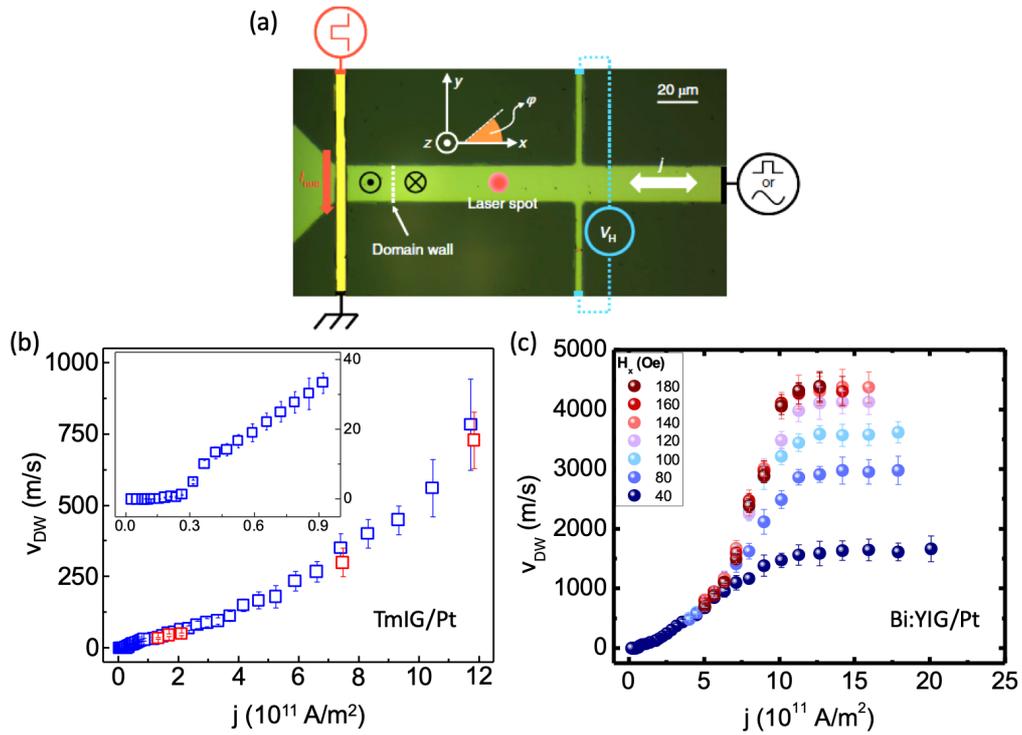

**Figure 9 – (a)** (Color online) A typical DW device with integrated Hall arms. The magnetization can be probed either by a laser spot, as shown, or magnetic imaging by using magneto-optic Kerr effect. DW velocity measurements as a function of the current density in TmIG/Pt **(b)** and in Bi:YIG/Pt **(c).** (Adapted from Refs. 98 and 105)

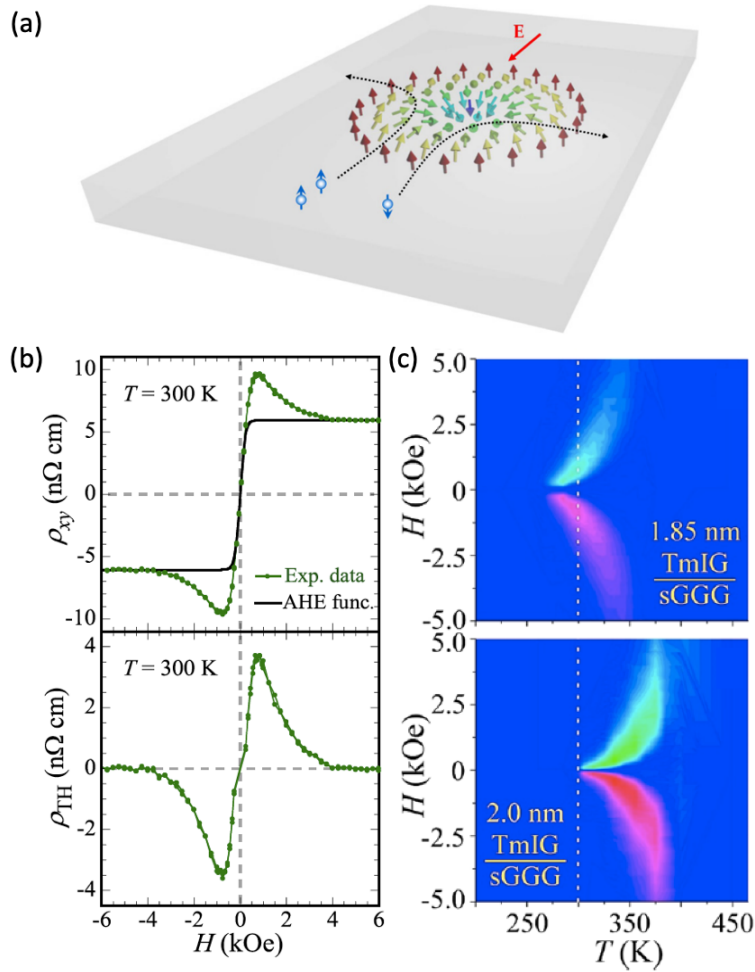

**Figure 10** – (Color online) **(a)** Schematics of a magnetic sykrmion and the emergence of the topological Hall effect due to polarization-dependent electron deflection. **(b)** The transverse Hall resistivity in TmIG(1.85 nm)/Pt at room temperature as a function of an out-of-plane field sweep and the extracted topological Hall effect signal (bottom panel). **(c)** Colored phase diagram showing the temperature and field dependence of the emergent topological Hall signal in layers with different TmIG thickness. (Adapted from Refs. 110 and 111)